\documentclass[reprint,twocolumn,longbibliography,
showpacs,preprintnumbers,
notitlepage,amsmath,amssymb,
prl,
]{revtex4-1}

\usepackage{graphicx}
\usepackage{dcolumn}
\usepackage{bm}
\usepackage{bbm}
\usepackage{diagbox}

\usepackage{graphicx}
\usepackage{enumerate}
\usepackage{subfigure}
\usepackage{epstopdf}
\usepackage[colorlinks,
linkcolor=red,
anchorcolor=black,
citecolor=blue]{hyperref}
\usepackage{amssymb}
\usepackage{framed}
\usepackage{tabularx}
\usepackage{booktabs}
\usepackage{float}
\usepackage[table]{xcolor}
\usepackage{array}
\usepackage{tikz}
\hypersetup{colorlinks=true}

\begin{document}
\title{Crystalline equivalent  boundary-bulk correspondence of two-dimensional topological phases
}
\author{Jian-Hao Zhang}
\email{jianhaozhang11@cuhk.edu.hk}
\affiliation{Department of Physics, The Chinese University of Hong Kong, Shatin, New Territories, Hong Kong, China}
\author{Shang-Qiang Ning}
\email{sqning91@gmail.com}
\affiliation{Department of Physics, The Chinese University of Hong Kong, Shatin, New Territories, Hong Kong, China}

\begin{abstract}
The boundary of topological phases of matter can manifest its topology nature, which leads to the so-called boundary-bulk correspondence (BBC) of topological phases. 
In this Letter, we construct a one-to-one correspondence between the boundary theories of fermionic SPT (fSPT) phases protected by crystalline symmetry and on-site symmetry in 2D fermionic systems, which follow the so-called crystalline equivalence principle. We dub such correspondence  \textit{crystalline equivalent BBC}. We illustrate this correspondence by two simple examples and as an application, we discover the topological boundary theory of 2D fSPT phase with spin-1/2 fermions, protected by a non-Abelian group $\mathbb{Z}_4\rtimes\mathbb{Z}_2^T$ with $A^4=\mathcal{T}^2=P_f$ where $A$ is the generator of $\mathbb{Z}_4$, from its crystalline equivalent partner---2D higher-order fSPT phase with spinless fermions, protected by $D_4$ symmetry. 
\end{abstract}

\newcommand{\lra}{\longrightarrow}
\newcommand{\xra}{\xrightarrow}
\newcommand{\ra}{\rightarrow}
\newcommand{\bs}{\boldsymbol}
\newcommand{\ul}{\underline}
\newcommand{\1}{\text{\uppercase\expandafter{\romannumeral1}}}
\newcommand{\2}{\text{\uppercase\expandafter{\romannumeral2}}}
\newcommand{\3}{\text{\uppercase\expandafter{\romannumeral3}}}
\newcommand{\4}{\text{\uppercase\expandafter{\romannumeral4}}}
\newcommand{\5}{\text{\uppercase\expandafter{\romannumeral5}}}
\newcommand{\6}{\text{\uppercase\expandafter{\romannumeral6}}}

\maketitle
\textit{Introduction} -- The interplay between symmetry and topology plays a central role in  the topological phases of quantum matter in recent years. In particular, symmetry protected topological (SPT) phases has been systematically constructed and classified by group cohomology (for bosonic), group supercohomology (for fermionic), spin cobordism or by gauging the corresponding global symmetry \cite{ZCGu2009,chen11a,XieChenScience,cohomology,E8,Lu12,invertible2,invertible3,special,general1,general2,Kapustin2014,Kapustin2015,Kapustin2017,Gu-Levin,gauging1,gauging3,dimensionalreduction,gauging2,2DFSPT,braiding}, which are beyond Landau symmetry-breaking paradigm. The best known example of SPT phases is topological insulator, which is protected by time-reversal and charge-conservation symmetry~\cite{KaneRMP,ZhangRMP}. 

Recently, SPT phases with crystalline symmetry have been intensively studied \cite{TCI,Fu2012,ITCI,reduction,building,correspondence,SET,230,BCSPT,Jiang2017,Kane2017,Shiozaki2018,ZDSong2018,defect,realspace,KenX,rotation,LuX,YMLu2018,Cheng2018,Hermele2018,Po2020,Huang2020PRR,Huang2021PRR,wallpaper,PEPS} because of  not only the conceptual interests, but also providing great opportunities for experimental realization~\cite{TCIrealization1,TCIrealization2,TCIrealization3,TCIrealization4}. In particular, a one-to-one correspondence between SPT phases with crystalline symmetry and on-site symmetry which is called ``\textit{crystalline equivalence principle}'' was proved rigorously in Refs. \cite{correspondence,defect}, and justified in 2D interacting fermionic systems with a twist of spin of fermions where spinless (spin-1/2) fermions should be mapped into spin-1/2 (spinless) fermions, by an explicit lower-dimensional block-state constructions of 2D crystalline fSPT phases, which is called \textit{topological crystals} \cite{dihedral,wallpaper}.

One celebrating phenomenon of the SPT phases is the boundary-bulk correspondence (BBC), namely the boundary of SPT phase must carry the corresponding symmetry anomaly, which manifests the nontrivial topology of the bulk phase. Usually the boundary of 2D SPT can either admit conformal field theory on its boundary  with certain 't Hooft anomaly \cite{Lu12,Ning21a}.
 The  crystalline or higher-order SPT phases can also have the BBC. Different from SPT with on-site symmetry, the boundaries of 2D crystalline SPT are usually almost gapped but with protected \textit{corner} zero modes \cite{Wang2018,Yan2018,Nori2018,Wangyuxuan2018,Ryu2018,Zhang2019}.  At  first glance, there is no direct relation of these two kinds of boundaries, even though their bulks obey the crystalline equivalence principle. However, we will show that the boundaries of the crystalline equivalent bulk topological phases can also follow crystalline equivalence principle. One key observation  is that the  boundary of SPT can  spontaneously break symmetry,  whose  domain wall however can trap nontrivial zero modes  that also manifest the nontrivial topology of the bulk. In light of the fact that their bulk are crystalline equivalent, the zero modes on the domain wall and the corner zero modes should be equivalent in some sense, which leads us to propose the so-called \emph{crystalline equivalent} BBC. 

In this Letter, we will demonstrate the crystalline equivalent BBC by constructing a one-to-one correspondence between the boundaries  of 2D fSPT phases protected by on-site symmetry and 2D higher-order fSPT phases with point group symmetry. The correspondence can be established  by treating the corner modes of 2D higher-order fSPT phases as ``crystalline symmetric" domain walls of 1D modes. For illustrating examples, we build the correspondence between reflection-symmetric topological superconductor (TSC) and the time reversal symmetric TSC  with $\mathcal{T}^2=-1$,  and also between $C_2$-symmetric TSC and  TSC protected by unitary on-site $\mathbb{Z}_2$ \cite{2DTSC,shinsei12,Gu-Levin,Qi12, Yao13,Gu-Levin}, whose bulks follow the  crystalline equivalence principle.
 Furthermore, as an application of the crystalline equivalent BBC, we construct  the boundary theory of one intrinsically interacting fSPT protected by non-Abelian $\mathbb{Z}_4\rtimes\mathbb{Z}_2^T$ symmetry through its crystalline partner $D_4$-symmetric TSC.
    The crystalline equivalent BBC we propose in this Letter would be extremely powerful for investigating the topological edge theory of fSPT phases with non-Abelian/antiunitary symmetry/mixed on-site and spatial symmetry.

\textit{Spinless fermion with reflection symmetry} -- Firstly we study the simplest case of spinless fermions with reflection symmetry. From topological crystals and explicit model construction, we know that for a 2D reflection-symmetric system with spinless fermions, there is a nontrivial higher-order topological phase with two Majorana corner modes $\xi_1$ and $\xi_2$, see Fig. \ref{higher order}. 

We provide an alternative comprehension of these Majorana corner modes: For $\xi_1$, consider two branches of itinerating Majorana modes $\gamma_\uparrow$ and $\gamma_\downarrow$ ($\uparrow$ and $\downarrow$ are virtual indices) on the boundary that move oppositely, with a mass term $m(x)$:
\begin{align}
H=\int\mathrm{d}x\cdot\gamma^T\left[i\sigma^3\partial_x+m(x)\sigma^2\right]\gamma
\label{eq:refl_mass}
\end{align}
where $\gamma(x)=\left(\gamma_\uparrow(x),\gamma_\downarrow(x)\right)^T$. The reflection symmetry $\bs{M}$ is defined as:
\begin{align}
\bs{M}:~
\gamma_\uparrow(x)\leftrightarrow\gamma_\downarrow(-x)
\label{M properties}
\end{align}
It is easy to verify that $H$ is invariant under $\bs{M}$ if $m(-x)=-m(x)$, which means the mass $m(x)$ has a domain-wall structure. For simplicity,  take $m(x)\sim x$, then the Hamiltonian $H$ reduces to:
\begin{align}
H=\int\mathrm{d}x\cdot\gamma^T\mathcal{H}(x)\gamma,~~\mathcal{H}(x)=i\sigma^3\partial_x+x\sigma^2
\end{align}
There is a Majorana zero mode localized at $x=0$ as a Gaussian wavepacket \cite{supplementary}: 
\begin{align}
|0\rangle= \mathcal{A} e^{-x^2/2}\left(1,1\right)^T
\label{M zero mode}
\end{align}
where $\mathcal{A}$ is a normalization factor. Equivalently, the Majorana corner modes of nontrivial higher-order topological phase in 2D reflection-symmetric system with spinless fermions can be treated as a domain-wall at the corner of the system.  In particular, this domain wall is  reflection symmetric.  To see this, denote $H_m(x)= 2 i  m(x) \gamma_\uparrow(x) \gamma_\downarrow(x)$, under $M$, $H_m(x)$ maps to $H_m(-x)$. The whole domain wall is symmetric under reflection. In other words, the domain wall is carrying neutral reflcetion quantum number.

\begin{figure}
\begin{tikzpicture}
\tikzstyle{sergio}=[rectangle,draw=none]
\draw[draw=red, thick] (-0.5,2.5)--(1.5,2.5)--(1.5,0.5)--(-0.5,0.5)--cycle;
\filldraw[fill=black, draw=black] (0.5,2.5)circle (2.5pt);
\filldraw[fill=black, draw=black] (0.5,0.5)circle (2.5pt);
\path (0.2222,2.8029) node [style=sergio] {$\xi_1$};
\path (0.2222,0.2193) node [style=sergio] {$\xi_2$};
\draw[thick,dashed] (0.5,-0.1) -- (0.5,3.1);
\path (0.7222,2.8029) node [style=sergio] {$M$};
\path (2.4,1.5) node [style=sergio] {$\Rightarrow$};
\draw[thick,color=red] (3,1.5) -- (6,1.5);
\draw[thick,dashed] (4.5,0.75) -- (4.5,2.25);
\path (5,2.5) node [style=sergio] {$M$};
\filldraw[fill=black, draw=black] (4.5,1.5)circle (2.5pt);
\path (4.215,1.1544) node [style=sergio] {$\xi_1$};
\end{tikzpicture}
\caption{Edge modes of 2D reflection-symmetric system with spinless fermions. Right panel is the zoom-in of the Majorana corner modes $\xi_1$. Reflection axis is depicted by dashed line.}
\label{higher order}
\end{figure}
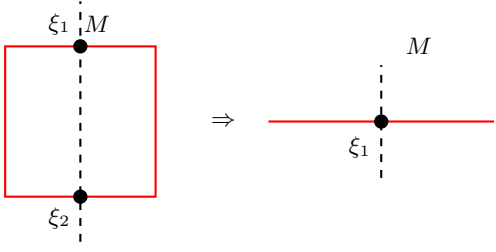

Subsequently, we define a effective ``time-reversal symmetry'' $\mathcal{T}$ in $\gamma$-basis and treat $\uparrow$ and $\downarrow$ as effective ``spin indices'': $\mathcal{T}=i\sigma^2 K,~~\mathcal{T}^2=-1$, where $K$ is the complex conjugate operator. It is easy to verify that the kinetic term is symmetric under $\mathcal{T}$:
\begin{align}
\mathcal{T}^{-1}(i\sigma^3\partial_x)\mathcal{T}=i\sigma^3\partial_x
\label{H0T}
\end{align}
But the mass term breaks $\mathcal{T}$:
\begin{align}
\mathcal{T}^{-1}\left[m(x)\sigma^2\right]\mathcal{T}=-m(x)\sigma^2
\end{align}
i.e., the domain wall structure is time reversal broken.

Furthermore, we investigate the symmetry properties of zero mode (\ref{M zero mode}) under reflection symmetry and ``time-reversal symmetry''. According to Eq. (\ref{M properties}), it is easy to see that this zero mode is reflection symmetric. On the other hand, under $\mathcal{T}$, the zero mode (\ref{M zero mode}) transforms as:
\begin{align}
\mathcal{T}|0\rangle=(i\sigma^2K)|0\rangle=\mathcal{A} e^{-x^2/2}\left(-1,1\right)^T
\end{align}
i.e., this zero mode breaks $\mathcal{T}$. However, this zero mode carries $\mathcal{T}^2=-1$.

To arrive at the helical edge theory of 2D time-reversal-invariant TSC \cite{2DTSC}, there are two ways:  one can turn off the time reversal broken domain wall, leaving only the kinetic term which is ``time-reversal-invariant'' [cf. Eq. (\ref{H0T})], or proliferate this domain wall that traps Majorana zero mode \cite{Max19}.
One can also go from the helical edge theory of 2D time-reversal-invariant TSC to obtain the Majorana corner zero modes of refelection symmetric TSC by adding the the reflection symmetric domain wall as in (\ref{eq:refl_mass}) and realizing reflection on helical majorana fermions as (\ref{M properties}).
This just establishes  the ``crystalline equivalence principle'' of BBC between time reversal and reflection symmetric TSC. 

\textit{Spin-1/2 fermion with $C_2$ symmetry} -- Repeatedly from topological crystals and explicit model construction, we know that for a 2-fold rotational-invariant 2D system with spin-1/2 fermions, there is a nontrivial higher-order topological phase with two Majorana corner modes $\xi_1$ and $\xi_2$ on the boundary of the system, see Fig. \ref{C2}. We introduce polar coordinates $(x,y)=r(\cos\theta,\sin\theta)$ to describe the Majorana corner modes at the north/south poles. Under the 2-fold rotation, the two zero modes $\xi_1$ and $\xi_2$ exchange.

\begin{figure}
\begin{tikzpicture}
\tikzstyle{sergio}=[rectangle,draw=none]
\draw[draw=red, thick] (-0.5,2.5)--(1.5,2.5)--(1.5,0.5)--(-0.5,0.5)--cycle;
\filldraw[fill=black, draw=black] (1.5,2.5)circle (2.5pt);
\filldraw[fill=black, draw=black] (-0.5,0.5)circle (2.5pt);
\path (1.2222,2.8029) node [style=sergio] {$\xi_1$};
\path (-0.1708,0.2193) node [style=sergio] {$\xi_2$};
\path (2.25,1.5) node [style=sergio] {$\Rightarrow$};
\filldraw[fill=green, draw=yellow] (0.5,1.5)circle (2.5pt);
\path (0.2521,1.2923) node [style=sergio] {$R$};
\draw[draw=red] (4.5,1.5)circle (30pt);
\filldraw[fill=black, draw=black] (4.5004,2.5492)circle (2.5pt);
\path (4.2774,2.226) node [style=sergio] {$\xi_1$};
\path (4.7258,0.73) node [style=sergio] {$\xi_2$};
\filldraw[fill=black, draw=black] (4.5083,0.4415)circle (2.5pt);
\path (4.2412,1.2286) node [style=sergio] {$R$};
\draw[->,thick] (3,1.5) -- (6,1.5);
\draw[->,thick] (4.5,0) -- (4.5,3);
\path (5.9228,1.2378) node [style=sergio] {$x$};
\path (4.7636,2.9067) node [style=sergio] {$y$};
\filldraw[fill=green, draw=yellow] (4.5,1.5)circle (2.5pt);
\end{tikzpicture}
\caption{Edge modes of 2D $C_2$-symmetric system with spin-1/2 fermions. Green dot represents the center of $C_2$.}
\label{C2}
\end{figure}
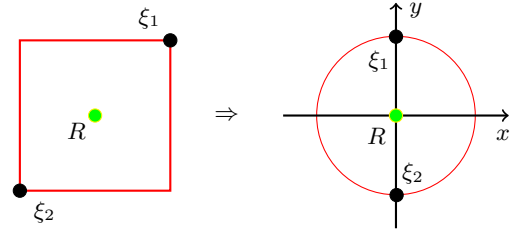

Consider two branches of itinerating Majorana modes $\gamma_A$ and $\gamma_B$ that move oppositely, with a mass term $m(\theta)\sim\cos\theta$:
\begin{align}
H=\int\mathrm{d}\theta\cdot\gamma^T\left(-i\sigma^3\partial_\theta+\cos\theta\cdot\sigma^2\right)\gamma
\label{eq:H_C2}
\end{align}
where $\gamma=(\gamma_A,\gamma_B)^T$, and the $C_2$ property of $\gamma$ is ($\mathrm{sgn}(x)=1$ for $x>0$, $\mathrm{sgn}(x)=-1$ for $x<0$):
\begin{align}
\bs{R}:~\gamma(\theta)\mapsto\mathrm{sgn}(\theta-\pi)\left(\gamma_A(\theta+\pi),\gamma_B(\theta+\pi)\right)^T
\label{C2 symmetry}
\end{align}
It is easy to verify that $H$ is $C_2$-symmetric. The mass term cos$(\theta)$ has a domain wall structure, which is $C_2$-symmetric.  Near northpole/southpole, the total Hamiltonian can be approximately expressed as:
\begin{align}
\begin{aligned}
&H^N=\int\mathrm{d}x\cdot\gamma^T\mathcal{H}^N\gamma,~~\mathcal{H}^N=i\sigma^3\partial_x+x\sigma^2\\
&H^S=\int\mathrm{d}x\cdot\gamma^T\mathcal{H}^S\gamma,~~\mathcal{H}^S=-i\sigma^3\partial_x-x\sigma^2
\end{aligned}
\end{align}
We concentrate on the physics near northpole, and the physics near southpole can be obtained from northpole by a 2-fold rotation. The Hamiltonian ${H}^N$ has a zero mode \cite{supplementary}:
\begin{align}
|0\rangle=\mathcal{A} e^{-x^2/2}\left(1,1\right)^T
\label{C2 zero mode}
\end{align}
Equivalently, the Majorana corner mode $\xi_1$ of nontrivial higher-order fSPT phase in 2D $C_2$-symmetric system with spin-1/2 fermions can be treated as a domain-wall at the northpole of a spherical geometry as illustrated in Fig. \ref{C2}. Similar for $\xi_2$ at the southpole.  The $C_2$ symmetric domain wall structure promises that $\xi_1$ and $\xi_2$ exchange under $\bs{R}$.

Subsequently, we define an effective ``$\mathbb{Z}_2$ on-site symmetry'' $O$ in $\gamma$-basis: $O=\sigma^3$, the kinetic term is symmetric under $O$:
\begin{align}
O^{-1}(-i\sigma^3\partial_\theta)O=-i\sigma^3\partial_\theta
\label{Z2}
\end{align}
But the mass term breaks this ``$\mathbb{Z}_2$ symmetry'':
\begin{align}
O^{-1}\left(\cos\theta\cdot\sigma^2\right)O=-\cos\theta\cdot\sigma^2
\end{align}
Furthermore, under $O$, the zero mode (\ref{C2 symmetry}) transforms as:
\begin{align}
O|0\rangle=\sigma^3|0\rangle=\mathcal{A} e^{-x^2/2}\left(1,-1\right)^T
\end{align}
i.e., this zero mode breaks $O$, however it has ${O}^2=1$.

Similarly to time reversal TSC, we have two ways to obtain the gapless  majorana edge modes of $\mathbb{Z}_2$ TSC: turn off the $\mathbb{Z}_2$-broken mass term cos($\theta$) \cite{shinsei12,Gu-Levin,Qi12, Yao13}, or proliferate the domain wall that traps majorana zero modes \cite{Max19}. One can also begin with gapless  majorana edge theory of $\mathbb{Z}_2$ TSC to construct the boundary corner modes of $C_2$-symmetric TSC by adding  mass term as in (\ref{eq:H_C2}) and realizing the $C_2$ symmetry $\bs{R}$  as (\ref{C2 symmetry}).  Then the crystalline equivalent BBC between unitary $\mathbb{Z}_2$ TSC and  $C_2$ symmetric TSC is just established.

\textit{Spinless fermion with $D_4$ symmetry} -- Topological crystals and explicit model construction \cite{dihedral,crystallineLSMOH} show that for a 2D $D_4$-symmetric system with spinless fermions ($D_4$ is 4-fold dihedral symmetry $D_4=C_4\rtimes\mathbb{Z}_2^M$ with two generators $\bs{R}\in C_4$ as a 4-fold rotation and $\bs{M}_1\in\mathbb{Z}_2^M$ as a reflection), there is a nontrivial higher-order fSPT phase, with 8 localized Majorana corner modes $\xi_j$ and $\xi_j'$ ($j=1,2,3,4$), see Fig. \ref{D4} \cite{crystallineLSMOH}. Similar to the $C_2$-symmetric case, we introduce polar coordinates $(x,y)=r(\cos\theta,\sin\theta)$ to describe the Majorana corner modes at poles (northpole, southpole, westpole and eastpole). 

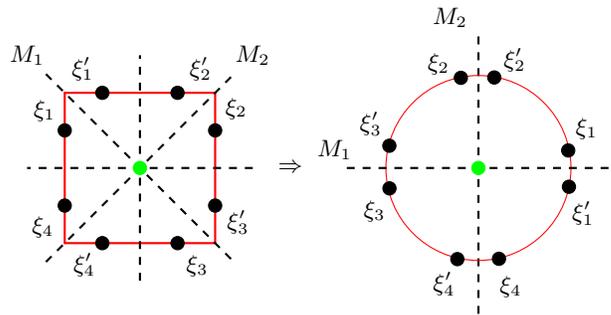
\begin{figure}
\begin{tikzpicture}
\tikzstyle{sergio}=[rectangle,draw=none]
\draw[draw=red, thick] (0,2.5)--(2,2.5)--(2,0.5)--(0,0.5)--cycle;
\filldraw[fill=black, draw=black] (1.5,2.5)circle (2.5pt);
\filldraw[fill=black, draw=black] (0.5,0.5)circle (2.5pt);
\path (-0.2687,2.258) node [style=sergio] {$\xi_1$};
\path (0.2383,2.8205) node [style=sergio] {$\xi_1'$};
\path (-0.5,3) node [style=sergio] {$M_1$};
\path (2.5,3) node [style=sergio] {$M_2$};
\filldraw[fill=black, draw=black] (2,2)circle (2.5pt);
\filldraw[fill=black, draw=black] (0,1)circle (2.5pt);
\filldraw[fill=black, draw=black] (1.5,0.5)circle (2.5pt);
\filldraw[fill=black, draw=black] (2,1)circle (2.5pt);
\filldraw[fill=black, draw=black] (0,2)circle (2.5pt);
\filldraw[fill=black, draw=black] (0.5,2.5)circle (2.5pt);
\path (2.2795,2.2918) node [style=sergio] {$\xi_2$};
\path (1.8051,2.8205) node [style=sergio] {$\xi_2'$};
\path (1.763,0.1964) node [style=sergio] {$\xi_3$};
\path (2.3038,0.7793) node [style=sergio] {$\xi_3'$};
\path (-0.2903,0.7142) node [style=sergio] {$\xi_4$};
\path (0.26,0.2181) node [style=sergio] {$\xi_4'$};
\draw[thick,dashed] (-0.25,0.25) -- (2.25,2.75);
\draw[thick,dashed] (1,0) -- (1,3);\draw[thick,dashed] (-0.25,2.75) -- (2.25,0.25);
\draw[thick,dashed] (2.5,1.5) -- (-0.5,1.5);
\path (3,1.5) node [style=sergio] {$\Rightarrow$};
\draw[thick,dashed] (5.5,3.25) -- (5.5,-0.5);
\draw[draw=red] (5.5,1.5)circle (35pt);
\path (5.1229,3.5) node [style=sergio] {$M_2$};
\filldraw[fill=black, draw=black] (5.2663,2.6995)circle (2.5pt);
\filldraw[fill=black, draw=black] (5.7105,2.704)circle (2.5pt);
\path (4.9706,2.8841) node [style=sergio] {$\xi_2$};
\path (5.9812,2.9432) node [style=sergio] {$\xi_2'$};
\draw[thick,dashed] (3.75,1.5) -- (7.25,1.5);
\filldraw[fill=black, draw=black] (6.6933,1.7333)circle (2.5pt);
\filldraw[fill=black, draw=black] (6.7032,1.2504)circle (2.5pt);
\path (6.8904,0.8982) node [style=sergio] {$\xi_1'$};
\path (6.9396,2.0218) node [style=sergio] {$\xi_1$};
\path (3.5871,1.7461) node [style=sergio] {$M_1$};
\filldraw[fill=black, draw=black] (4.3157,1.7959)circle (2.5pt);
\filldraw[fill=black, draw=black] (4.3173,1.2287)circle (2.5pt);
\path (4.1186,0.9259) node [style=sergio] {$\xi_3$};
\path (4.0693,2.0889) node [style=sergio] {$\xi_3'$};
\filldraw[fill=black, draw=black] (5.7732,0.2916)circle (2.5pt);
\filldraw[fill=black, draw=black] (5.2197,0.2873)circle (2.5pt);
\path (5.9293,-0.0801) node [style=sergio] {$\xi_4$};
\path (5.0438,-0.0408) node [style=sergio] {$\xi_4'$};
\filldraw[fill=green, draw=green] (5.5,1.5)circle (2.5pt);
\filldraw[fill=green, draw=green] (1,1.5)circle (2.5pt);
\end{tikzpicture}
\caption{Edge modes of 2D $D_4$-symmetric system with spinless fermions. All dashed lines are reflection axes and green dot is the center of 4-fold rotation symmetry $C_4$.}
\label{D4}
\end{figure}

We introduce 4 branches of itinerating Majorana modes $\gamma=(\gamma_1^\uparrow,\gamma_1^\downarrow,\gamma_2^\uparrow,\gamma_2^\downarrow)^T$ on the boundary, where the Majorana modes with $\uparrow$ and $\downarrow$ indices move in opposite directions:
\begin{align}
H_0=\int\mathrm{d}\theta\cdot\gamma^T\left[i(\tau^3\otimes\sigma^3)\partial_\theta\right]\gamma
\end{align}
where $\tau^{1,2,3}$ are Pauli matrices characterizing the pseudo-spin indices. These Majorana modes have the following symmetry properties:
\begin{align}
\begin{aligned}
\bs{M}_1:~&\left\{
\begin{aligned}
&\theta\mapsto2\pi-\theta\\
&\left(\gamma_\uparrow^1,\gamma_\downarrow^1,\gamma_\uparrow^2,\gamma_\downarrow^2\right)\mapsto\left(\gamma_\uparrow^2,\gamma_\downarrow^2,\gamma_\uparrow^1,\gamma_\downarrow^1\right)
\end{aligned}
\right.\\
\bs{R}:~&\left\{
\begin{aligned}
&\theta\mapsto\theta+\pi/2\\
&\left(\gamma_\uparrow^1,\gamma_\downarrow^1,\gamma_\uparrow^2,\gamma_\downarrow^2\right)\mapsto\left(\gamma_\uparrow^1,\gamma_\downarrow^1,\gamma_\uparrow^2,\gamma_\downarrow^2\right)
\end{aligned}
\right.
\end{aligned}
\label{D4 symmetry}
\end{align}
That satisfy $\bs{R}^4=\bs{M}_1^2=1$. It is easy to verify that $H_0$ is invariant under $D_4$ generators (\ref{D4 symmetry}). We further consider a $D_4$-symmetric ``mass term'' with a spatial-dependent mass $m_j(\theta)$:
\begin{align}
H_m=\int\mathrm{d}\theta\cdot\gamma^T\left[\sin(2\theta)\cdot(\tau^3\otimes\sigma^2)\right]\gamma
\end{align}
And we can straightforwardly confirm that $H_m$ is invariant under $D_4$ generators (\ref{D4 symmetry}). After investigating the symmetry properties of the total Hamiltonian $H=H_0+H_m$, we can concentrate on the Hamiltonian near each pole (intersection between the boundary of system and reflection axes along the vertical/horizontal directions, see right panel of Fig. \ref{D4}). Near eastpole, from $\theta\sim0$ we obtain $\sin(2\theta)\sim0$, the low-energy physics near the eastpole can be described by the following Hamiltonians:
\begin{align}
H^E=\int\mathrm{d}y\cdot\gamma^T\left[i(\tau^3\otimes\sigma^3)\partial_y+y(\tau^3\otimes\sigma^2)\right]\gamma
\end{align}
The Hamiltonian $H^E$ has two zero modes ($x=r$):
\begin{align}
\begin{aligned}
&|0\rangle_{1}=\mathcal{A}e^{-y^2/2}\left(1,1,0,0\right)^T\\
&|0\rangle_{1'}=\mathcal{A}e^{-y^2/2}\left(0,0,1,1\right)^T
\end{aligned}
\label{D4 zero modes east}
\end{align}
Equivalently, the Majorana corner modes $\gamma_1$ and $\gamma_1'$ of nontrivial higher-order topological phase in 2D $D_4$-symmetric systems with spinless fermions can be treated as two domain walls near the eastpole of the spherical geometry in Fig. \ref{D4}. Similar for Majorana corner modes at other poles with $\theta\sim\pi/2,\pi,3\pi/2$ at which $\sin(2\theta)\sim0$. Near westpole ($x=-r$):
\begin{align}
\begin{aligned}
&|0\rangle_{3}=\mathcal{A}e^{-y^2/2}\left(1,1,0,0\right)^T\\
&|0\rangle_{3'}=\mathcal{A}e^{-y^2/2}\left(0,0,1,1\right)^T
\end{aligned}
\label{D4 zero modes west}
\end{align}
Near north and south poles ($y=\pm r$):
\begin{align}
\begin{aligned}
&|0\rangle_{2,4}=\mathcal{A}e^{-x^2/2}\left(1,1,0,0\right)^T\\
&|0\rangle_{2',4'}=\mathcal{A}e^{-x^2/2}\left(0,0,1,1\right)^T
\end{aligned}
\label{D4 zero modes north and south}
\end{align}

Subsequently we define an effective ``$\mathbb{Z}_4$ on-site symmetry'' $A$ (with $A^4=-1$):
\begin{align}
A=\frac{1}{\sqrt{2}}\left(
\begin{array}{ccc}
\mathbbm{1}_{2\times2} & -i\sigma^2\\
-i\sigma^2 & \mathbbm{1}_{2\times2}
\end{array}
\right)
\label{Z4 symmetry}
\end{align}
$H_0$ is symmetric under $A$, mass term $H_m$ breaks $A$:
\begin{align}
\begin{aligned}
&A^{-1}i(\tau^3\otimes\sigma^3)A=i(\tau^3\otimes\sigma^3)\\
&A^{-1}(\tau^3\otimes\sigma^2)A\ne\tau^3\otimes\sigma^2
\end{aligned}
\end{align}
Then we define an effective ``time-reversal symmetry'' $\mathcal{T}$:
\begin{align}
\mathcal{T}=i(\tau^2\otimes\mathbbm{1}_{2\times2})K,~~\mathcal{T}^2=-1
\label{time reversal symmetry}
\end{align}
It is easy to verify that $H_0$ and $H_m$ preserve $\mathcal{T}$:
\begin{align}
\begin{aligned}
&\mathcal{T}^{-1}i(\tau^3\otimes\sigma^3)\mathcal{T}=i(\tau^3\otimes\sigma^3)\\
&\mathcal{T}^{-1}(\tau^3\otimes\sigma^2)\mathcal{T}=\tau^3\otimes\sigma^2
\end{aligned}
\end{align}
All zero modes [cf. Eqs. (\ref{D4 zero modes east})-(\ref{D4 zero modes north and south})] are symmetric under $D_4$  and also $\mathcal{T}$ symmetry, but not symmetric under $A$ \cite{supplementary}.

Nevertheless, if we ``release'' the Majorana corner modes $\xi_j$ and $\xi_j'$ (by removing the domain walls, $j=1,2,3,4$), the Hamiltonian $H$ reduces to $H_0$ which is ``$\mathbb{Z}_4$-symmetric'' [cf. Eq. (\ref{Z4 symmetry})] and ``time-reversal symmetric'' [cf. Eq. (\ref{time reversal symmetry})]. Hence the Hamiltonian $H_0$ describes the edge theory of 2D $(\mathbb{Z}_4\rtimes\mathbb{Z}_2^T)$-symmetric systems with spin-1/2 fermions. Furthermore, we study if these edge modes can not be gapped in a symmetric way. We bosonize the edge theory $H_0$ in terms of $\gamma_\sigma^{1,2}$ ($\sigma=\uparrow,\downarrow$):
\begin{align}
e^{i\phi_{1}}=\gamma_1^\uparrow+i\gamma_2^\downarrow,~~~e^{i\phi_{2}}=\gamma_1^\downarrow+i\gamma_2^\uparrow
\end{align}
And the topological edge theory $H_0$ can be rephrased in terms of bosonic fields $\Phi=(\phi_{1},\phi_{2})^T$:
\begin{align}
\mathcal{L}_{\mathrm{edge}}=\frac{K_{IJ}}{4\pi}(\partial_x\Phi^I)(\partial_t\Phi^J)+\frac{V_{IJ}}{8\pi}(\partial_x\Phi^I)(\partial_x\Phi^J)
\end{align}
where $K=\sigma^z$ as the $K$-matrix characterizing the topology of $\mathcal{L}_{\mathrm{edge}}$. Under $A$ and $\mathcal{T}$, the field of edge modes $\Phi$ transforms as \cite{Ning21a}:
\begin{align}
A:\Phi\mapsto W^{A}\Phi+\delta\Phi^{A},~\mathcal{T}:\Phi\mapsto W^{\mathcal{T}}\Phi+\delta\Phi^{\mathcal{T}}
\end{align}
where
\begin{align}
\left\{
\begin{aligned}
&W^A=\mathbbm{1}_{2\times2}\\
&W^{\mathcal{T}}=\sigma^x
\end{aligned}
\right.,~~\left\{
\begin{aligned}
&\delta\Phi^A={\pi}(1/4,-1/4)^T\\
&\delta\Phi^{\mathcal{T}}={\pi}(1/2,1/2)^T
\end{aligned}
\right.
\label{symmetry}
\end{align}
We now try to construct interaction terms that gap out the edge without breaking the $A$ and $\mathcal{T}$ symmetries, either explicitly or spontaneously. Consider the backscattering terms of the form:
\begin{align}
U=\sum\limits_jU(\Lambda_j)=\sum\limits_jU(x)\cos\left[\Lambda_j^TK\Phi-\alpha(x)\right]
\label{backscattering}
\end{align}
The backscattering term (\ref{backscattering}) can gap out the edge as long as the vectors $\{\Lambda_j\}$ satisfy the ``null-vector'' conditions \cite{Haldane1995} for $\forall i,j$:
\begin{align}
\Lambda_i^TK\Lambda_j=0
\end{align}
The simplest term is $\Lambda_1=(1,1)$ or $\Lambda_2=(1,-1)$.  However, $\Lambda_1$ breaks $\mathcal{T}$ symmetry and $\Lambda_2$ break $A$ symmetry.   We turn to the next simplest term $\Lambda_3=(2,2)$ or $\Lambda_4=(2,-2)$. $\Lambda_3$ preserve all the symmetry but leads to spontaneously symmetry breaking, i.e., $\langle\phi_-\rangle$ where $\phi_-:=\phi_1-\phi_2$, has two energy vacca: $0$ and $\pi$ (take $\alpha(x)=0$ for simplicity) which transform into each other by $A^2$. Similar analysis show  $\Lambda_4$ would spontaneously break $\mathcal{T}$. So it seems to be no way to symmetrically gap out the edge fields. One may guess it is possible to stack trivial edge fields to gap them out together symmetrically. We argue that in fact it is impossible by gauging the fermion parity symmetry. It turns out that the fermion parity flux carries projective representation of $\mathbbm{Z}_4\rtimes \mathbbm{Z}_2^{T}$ \cite{supplementary}.
 Equivalently, $H_0$ or $\mathcal{L}_{\mathrm{edge}}$ with (\ref{symmetry}) characterizes the nontrivial topological edge theory for 2D $(\mathbb{Z}_4\rtimes\mathbb{Z}_2^T)$-symmetric fSPT phase with spin-1/2 fermions.

\textit{Conclusion and discussion} -- Crystalline equivalence principle and boundary-bulk correspondence (BBC) are two celebrated aspects in topological phases of matter. In the 
Letter, we build a bridge between these two phenomena, by proposing the so-called \textit{crystalline equivalent BBC}. We demonstrate the idea of  crystalline equivalent BBC by studying some familar exmaples: 2D time reversal  vs reflection TSC,  and 2D unitary $\mathbbm{Z}_2$  vs $C_2$ rotational TSC,  which are both  crystalline equivalent. As a nontrivial application, we construct the edge theory of the interacting fSPT protected by $\mathbbm{Z}_4\rtimes \mathbbm{Z}_2^{T}$ with $A^4=\mathcal{T}=P_f$ from its crystalline partners $D_4$ higher order TSC. Generally speaking, the single majorana zero modes on the domain wall/ the projective representation of fermion parity flux are related to the $n_1/n_2$ data in the algebraic description of fSPT.  Constructing the boundary of SPT phase from translational topological phase (a special crystalline topological phases) is discussed in Refs. \cite{Max18,Meng18fLSM,JosephMeng19}. However, the translational topological phases and the corresponding SPT are not crystalline equivalent.  Moreover,
the way that we treat the corner modes as crystalline  ``domain wall" is fundamentally different from the way by utilizing the translational symmetry as in \cite{Max18,Meng18fLSM,JosephMeng19}.  The proposed crystalline equivalent BBC may shed light on the preparing the SPT states by quantum circuit in quamtum similator or quantum processor. It would be very important in studying the BBC of interacting SPT, especially for non-Abelian/antiunitary symmetry group, and  furthermore can be generalized to bosonic systems, 3D systems and the SPT systems jointly protected by crystalline symmetry $SG$ and on-site symmetry $G_0$, which are left for future study.

\textit{Acknowledgements} --   We thank Z.C Gu, Z. Bi and Z.X Liu for enlightening discussions.This work is supported by Direct Grant No. 4053409 from The Chinese University of Hong Kong and funding from Hong Kong's Research Grants Council (GRF No.14306918, ANR/RGC Joint Research Scheme No. A-CUHK402/18).


\providecommand{\noopsort}[1]{}\providecommand{\singleletter}[1]{#1}%
%


\pagebreak

\clearpage

\appendix
\setcounter{equation}{0}
\newpage

\renewcommand{\thesection}{S-\arabic{section}} \renewcommand{\theequation}{S%
\arabic{equation}} \setcounter{equation}{0} \renewcommand{\thefigure}{S%
\arabic{figure}} \setcounter{figure}{0}

\centerline{\textbf{Supplemental Material}}

\maketitle

\section{Majorana zero modes as a domain wall}
In the main text, we have concluded that for a 2D higher-order fSPT phase protected by reflection symmetry $\bs{M}$ with spinless fermions, the ``domain-wall'' physics near $x=0$ is described by the Hamiltonian:
\begin{align}
H=\int\mathrm{d}x\cdot\gamma^T\mathcal{H}(x)\gamma
\end{align}
where $\gamma=\left(\gamma_\uparrow,\gamma_\downarrow\right)^T$ and
\begin{align}
\mathcal{H}(x)=i\sigma^3\partial_x+x\sigma^2
\end{align}
To get the zero-energy solution, we define an alternative basis from a unitary transformation on $\gamma$:
\begin{align}
\chi=\left(
\begin{array}{ccc}
\chi_1\\
\chi_2
\end{array}
\right)=\frac{1}{\sqrt{2}}(\sigma^1+\sigma^3)\gamma=\frac{1}{\sqrt{2}}\left(
\begin{array}{ccc}
\gamma_\uparrow+\gamma_\downarrow\\
\gamma_\uparrow-\gamma_\downarrow
\end{array}
\right)
\end{align}
Under the basis $\chi$, the Hamiltonian $\mathcal{H}$ will be transformed to:
\begin{align}
\mathcal{H}'&=\frac{1}{\sqrt{2}}(\sigma^1+\sigma^3)\big[i\sigma^3\partial_x+x\sigma^2\big]\frac{1}{\sqrt{2}}(\sigma^1+\sigma^3)\nonumber\\
&=i\sigma^1\partial_x-x\sigma^2
\end{align}
Define the effective creation/annihilation operators $a$ and $a^\dag$ in terms of $x$ and $\partial_x$:
\begin{align}
\left\{
\begin{aligned}
&a=\frac{1}{\sqrt{2}}(x+\sigma^3\partial_x)\\
&a^\dag=\frac{1}{\sqrt{2}}(x-\sigma^3\partial_x)
\end{aligned}
\right.
\end{align}
with commutation relation:
\[
\left[a,a^\dag\right]=\frac{1}{2}\left[x+\sigma^3\partial_x,x-\sigma^3\partial_x\right]=\sigma^3
\]
Then we can rephrase the Hamiltonian $\mathcal{H}'^2$ in terms of $a$ and $a^\dag$ we defined above:
\begin{align}
\mathcal{H}'^2=(i\sigma^1\partial_x-x\sigma^2)^2=-\partial_x^2+x^2+\sigma^3=2a^\dag a
\end{align}
So if $\mathcal{H}'^2$ has a zero mode, so do $\mathcal{H}'$. Suppose $|0\rangle$ is a zero mode of $\mathcal{H}'^2$ that is proportional to $(1,0)^T$ and satisfying $a|0\rangle=0$ in $\chi$-basis:
\begin{align}
a|0\rangle=(x+\partial_x)|0\rangle,~\Rightarrow~|0\rangle\propto e^{-x^2/2}\left(
\begin{array}{ccc}
1\\
0
\end{array}
\right)
\end{align}
that is a Gaussian wavepacket localized near $x=0$. As the consequence, $\mathcal{H}'^2$ (and thus $\mathcal{H}'$) has zero mode $|0\rangle$ that is localized near $x=0$. In $\gamma$-basis, this zero mode is expressed as:
\begin{align}
|0\rangle\propto e^{-x^2/2}\frac{1}{\sqrt{2}}\left(
\begin{array}{ccc}
1\\
1
\end{array}
\right)
\label{M zero modeS}
\end{align}

\section{Symmetry properties of Majorana corner modes in $D_4$-symmetric case}
In the main text, for 2D $D_4$-symmetric systems with spinless fermions, we have reformulated the Majorana corner modes of the corresponding higher-order fSPT phase in terms of the domain walls on the boundary. In particular, these domain walls can be expressed in terms of the following basis:
\[
\gamma=\left(\gamma_\uparrow^1,\gamma_\downarrow^1,\gamma_\uparrow^2,\gamma_\downarrow^2\right)^T
\]
The Majorana corner modes at other poles can also be formulated in $\gamma$-basis:
\begin{align}
\begin{aligned}
&|0\rangle_{1,3}=\mathcal{A}e^{-y^2/2}\left(1,1,0,0\right)^T\\
&|0\rangle_{1',3'}=\mathcal{A}e^{-y^2/2}\left(0,0,1,1\right)^T\\
&|0\rangle_{2,4}=\mathcal{A}e^{-x^2/2}\left(1,1,0,0\right)^T\\
&|0\rangle_{2',4'}=\mathcal{A}e^{-x^2/2}\left(0,0,1,1\right)^T
\end{aligned}
\label{D4 zero modes north and southS}
\end{align}
Under 4-fold rotation $\bs{R}\in C_4$ and reflection $\bs{M}_1$, these zero modes transform as:
\begin{align}
&\bs{R}:~\left(|0\rangle_1,|0\rangle_{1'},|0\rangle_{2},|0\rangle_{2'},|0\rangle_{3},|0\rangle_{3'},|0\rangle_{4},|0\rangle_{4'}\right)\nonumber\\
&\mapsto\left(|0\rangle_{2},|0\rangle_{2'},|0\rangle_{3},|0\rangle_{3'},|0\rangle_{4},|0\rangle_{4'},|0\rangle_1,|0\rangle_{1'}\right)
\end{align}
and
\begin{align}
&\bs{M}_1:~\left(|0\rangle_1,|0\rangle_{1'},|0\rangle_{2},|0\rangle_{2'},|0\rangle_{3},|0\rangle_{3'},|0\rangle_{4},|0\rangle_{4'}\right)\nonumber\\
&\mapsto\left(|0\rangle_{1'},|0\rangle_{1},|0\rangle_{4'},|0\rangle_{4},|0\rangle_{3'},|0\rangle_{3},|0\rangle_{2'},|0\rangle_{4}\right)
\end{align}
i.e., all Majorana corner modes are $D_4$-invariant. Alternatively, we can phrase the $D_4$ symmetry properties in a more transparent way be redefine the Majorana zero modes:
\begin{align}
\begin{aligned}
&|z\rangle_{1,3}=(|0\rangle_{1,3}+|0\rangle_{1',3'})/\sqrt{2}\\
&|z\rangle_{1',3'}=(|0\rangle_{1,3}-|0\rangle_{1',3'})/\sqrt{2}\\
&|z\rangle_{2,4}=(|0\rangle_{2,4}+|0\rangle_{2',4'})/\sqrt{2}\\
&|z\rangle_{2',4'}=(|0\rangle_{2,4}-|0\rangle_{2',4'})/\sqrt{2}
\end{aligned}
\end{align}
Under 4-fold rotation $\bs{R}\in C_4$ and reflection $\bs{M}_1$, these zero modes transform as:
\begin{align}
&\bs{R}:~\left(|z\rangle_1,|z\rangle_{1'},|z\rangle_{2},|z\rangle_{2'},|z\rangle_{3},|z\rangle_{3'},|z\rangle_{4},|z\rangle_{4'}\right)\nonumber\\
&\mapsto\left(|z\rangle_{2},|z\rangle_{2'},|z\rangle_{3},|z\rangle_{3'},|z\rangle_{4},|z\rangle_{4'},|z\rangle_1,|z\rangle_{1'}\right)
\end{align}
and
\begin{align}
&\bs{M}_1:~\left(|z\rangle_1,|z\rangle_{1'},|z\rangle_{2},|z\rangle_{2'},|z\rangle_{3},|z\rangle_{3'},|z\rangle_{4},|z\rangle_{4'}\right)\nonumber\\
&\mapsto\left(|z\rangle_{1},-|z\rangle_{1'},|z\rangle_{4},-|z\rangle_{4'},|z\rangle_{3},-|z\rangle_{3'},|z\rangle_{2},-|z\rangle_{2'}\right)
\end{align}
i.e., Majorana zero modes $\left\{|z\rangle_j\big|j=1,2,3,4;1',2',3',4'\right\}$ carry charges of reflection generator $\bs{M}_1$.

In the main text, we have defined an effective ``$\mathbb{Z}_4$ on-site symmetry'' $A$ which satisfies $A^4=-1$. Under $A$, these Majorana corner modes will be transformed as:
\begin{align}
\begin{aligned}
&|0\rangle_{1,3}\mapsto\mathcal{A}e^{-y^2/2}\left(1,1,-1,1\right)^T\\
&|0\rangle_{1',3'}\mapsto\mathcal{A}e^{-y^2/2}\left(-1,1,1,1\right)^T\\
&|0\rangle_{2,4}\mapsto\mathcal{A}e^{-x^2/2}\left(1,1,-1,1\right)^T\\
&|0\rangle_{2',4'}\mapsto\mathcal{A}e^{-x^2/2}\left(-1,1,1,1\right)^T
\end{aligned}
\end{align}
i.e., these Majorana corner modes are not invariant under $A$ symmetry. We note that the symmetry $A$ is on-site and does not transform zero at one position to another. 
 Furthermore, we have defined another effective ``time-reversal symmetry'' $\mathcal{T}$ which satisfies $\mathcal{T}^2=-1$. Under $\mathcal{T}$, these Majorana corner modes will be transformed as:
\begin{align}
\begin{aligned}
&|0\rangle_{1,3}\mapsto\mathcal{A}e^{-y^2/2}\left(0,0,-1,-1\right)^T=-|0\rangle_{1',3'}\\
&|0\rangle_{1',3'}\mapsto\mathcal{A}e^{-y^2/2}\left(1,1,0,0\right)^T=|0\rangle_{1,3}\\
&|0\rangle_{2,4}\mapsto\mathcal{A}e^{-x^2/2}\left(0,0,-1,-1\right)^T=-|0\rangle_{2',4'}\\
&|0\rangle_{4}^{N,S}\mapsto\mathcal{A}e^{-x^2/2}\left(1,1,0,0\right)^T=|0\rangle_{2,4}
\end{aligned}
\end{align}
i.e., there Majorana corner modes are invariant under $\mathcal{T}$ symmetry. 

\section{Projective representaiton of fermion parity flux in gauged $\mathbb{Z}_4\rtimes\mathbb{Z}_2^T$ fermionic SPT}
In the main text, the edge theory $\mathbb{Z}_4\rtimes\mathbb{Z}_2^T$ fermionic SPT is given by $K=\sigma_z$ together with the symmetry realization Eq.29 in the main text, i.e.,
\begin{align}
\left\{
\begin{aligned}
&W^A=\mathbbm{1}_{2\times2}\\
&W^{\mathcal{T}}=\sigma^x
\end{aligned}
\right.,~~\left\{
\begin{aligned}
&\delta\Phi^A={\pi}(1/4,-1/4)^T\\
&\delta\Phi^{\mathcal{T}}={\pi}(1/2,1/2)^T
\end{aligned}
\right.
\label{app_symmetry}
\end{align}
Especially, the fermion parity symmetry is realized as 
\begin{align}
W^{P_f}=\mathbbm{1}_{2\times2}\,, \quad \delta\phi^{P_f}=(\pi, \pi)^T.
\label{app_fermion_parity}
\end{align}
Following the method in Ref. \cite{Ning21aS}, we can gauge the fermion parity symmetry and the fermion parity flux are represented by the ``fractionalized" vertext operators $e^{i \frac{\phi}{2}}$ where $\phi=\phi_1+\phi_2$. Now we study the symmetry properties of  the remaining symmetry $\mathbb{Z}_4\rtimes\mathbb{Z}_2^T$. In fact, under symmetry, the fermion parity flux form a doublet, which is represented by $(e^{i \frac{\phi}{2}}, e^{-i \frac{\phi}{2}})^T$. The components of the doublet differ by attaching  local fermions. Under $A$ and $\mathcal{T}$, 
\begin{align}
A&:\begin{pmatrix} e^{i \frac{\phi}{2}} \\ e^{-i \frac{\phi}{2}}\end{pmatrix}\rightarrow \begin{pmatrix} 1 &0 \\
0 &1\end{pmatrix}\begin{pmatrix} e^{i \frac{\phi}{2}} \\ e^{-i \frac{\phi}{2}}\end{pmatrix}\\
\mathcal{T}&:\begin{pmatrix} e^{i \frac{\phi}{2}} \\ e^{-i \frac{\phi}{2}}\end{pmatrix}\rightarrow \begin{pmatrix} 0 &i \\
-i &0\end{pmatrix}\begin{pmatrix} e^{i \frac{\phi}{2}} \\ e^{-i \frac{\phi}{2}}\end{pmatrix}
\end{align}
Namely, acting on the fermion parity flux doublet, $U_A=\mathbbm{1}_{2\times 2}$ and $U_{\mathcal{T}}=-\sigma_y K$. Recalling that the group relation $\mathcal{T}A\mathcal{T}^{-1}=A^{-1}$, namely $A$ and $\mathcal{T}$ do not commute, such a realization of $A,\mathcal{T}$ is indeed projective.  One can compute the  invariants for this projective representation, that is, 
\begin{align}
\begin{aligned}
&\mathcal{I}_1=n_2(A^2\mathcal{T},A^2\mathcal{T})=-1\\
&\mathcal{I}_2=n_2(A^3\mathcal{T},A^3\mathcal{T})=-1
\end{aligned}
\end{align}
where $n_2$ is the 2-cocycle corresponding to this projective representation \cite{yang2017irreducibleS}. We note that the corresponding 2-cocycle  with such invariants are in fact nontrivial in $\mathcal{H}^2\left[\mathbb{Z}_4\rtimes\mathbb{Z}_2^T, U(1)\right]$, and hence nontrivial  in $\mathcal{H}^2(\mathbb{Z}_4\rtimes\mathbb{Z}_2^T,\mathbb{Z}_2)$ which is the true symmetry fractionalization classification of fermion parity flux.  
In fact, the symmetry fractionalization class $(-1)^{n_2}$ of fermion parity flux  corresponds to the complex fermion decoration in the supercohomology theory for fermionic SPT \cite{JosephMeng19S}. In the classification of fermionic SPT, the complex fermion decoration with $n_2=w_2$ is trivialized \cite{general2S}. [The meaning of $w_2$ is that it defines the extension that characterizes the fermionic symmetry group, see Eq. (\ref{app_extension_fs})]
Such projective representation $U_A$ and $U_\mathcal{T}$, labeled by $n_2$, differs from the $\omega_2$ in Eq. (\ref{app_extension_fs}) since they have  different invariants [see Eq. (\ref{app_w2_inv})].  Therefore, the edge theory indeed corresponds to a nontrivial fermionic SPT. 

\section{Representation of $\mathbb{Z}_4\rtimes\mathbb{Z}_2^T$ in 2D system with spin-1/2 fermions}
In this section, we demonstrate that 4 Majorana fermions $\gamma_\sigma^j$ ($\sigma=\uparrow,\downarrow$ and $j=1,2$) introduced in the main text, with the following symmetry properties:
\begin{align}
A=\frac{1}{\sqrt{2}}\left(
\begin{array}{ccc}
\mathbbm{1}_{2\times2} & -i\sigma^2\\
-i\sigma^2 & \mathbbm{1}_{2\times2}
\end{array}
\right)
\label{Z4}
\end{align}
and
\begin{align}
\mathcal{T}=i(\tau^2\otimes\mathbbm{1}_{2\times2})K
\label{time reversal}
\end{align}
realize a reprensentation of $\mathbb{Z}_4\rtimes\mathbb{Z}_2^T$ group in 2D system with spin-1/2 fermions, where $A\in\mathbb{Z}_4$ and $\mathcal{T}\in\mathbb{Z}_2^T$ are two generators of the symmetry group $\mathbb{Z}_4\rtimes\mathbb{Z}_2^T$. For a fermionic system, there is always a fermion parity symmetry $\mathbb{Z}_2^f=\{1,P_f=(-1)^{F}\}$, where $F$ is the total number of fermions. The spin of fermions is characterized by the factor system $\omega_2$ of the following short exact sequence:
\begin{align}
0\rightarrow\mathbb{Z}_2^f\rightarrow G_f\rightarrow\mathbb{Z}_4\rtimes\mathbb{Z}_2^T\rightarrow0
\end{align}
where $G_f$ depicts the total symmetry group of the system, as a group extension of $\mathbb{Z}_4\rtimes\mathbb{Z}_2^T$ and fermion parity $\mathbb{Z}_2^f$. $\omega_2$ is an element of the following group 2-cohomology:
\begin{align}
\omega_2\in\mathcal{H}^2\left[\mathbb{Z}_4\rtimes\mathbb{Z}_2^T,\mathbb{Z}_2^f\right]=\mathbb{Z}_2^3
\end{align}
In particular, the spin-1/2 fermions corresponding to the 2-cocycle $\omega_2$ satisfying the following conditions:
\begin{align}
\left\{
\begin{aligned}
&A^4=P_f\\
&\mathcal{T}^2=P_f\\
&\mathcal{T}A\mathcal{T}^{-1}A=1
\end{aligned}
\right.
\label{spin-1/2}
\end{align}
To satisfy these conditions, we consider the 2-cocycle $\omega_2$ as following. For $\forall a_g,b_h\in\mathbb{Z}_4\rtimes\mathbb{Z}_2^T$ defined as:
\begin{align}
\mathbb{Z}_4\rtimes\mathbb{Z}_2^T=\left\{(a,g)=a_g\Big|0\leq a\leq3,0\leq g\leq1\right\}
\end{align}
we choose
\begin{align}
\omega_2(a_g,b_h)=&\left\lfloor\frac{\left[(-1)^{g+h}a\right]_{2n}+\left[(-1)^{h}b\right]_{2n}}{2n}\right\rfloor\nonumber\\
&+(1-\delta_a)(a+1)h+g\cdot h
\label{app_extension_fs}
\end{align}
where we define $[x]_n\equiv x(\mathrm{mod}~n)$ with $n=2$, $\lfloor x\rfloor$ as the greatest integer less than or equal to $x$, and
\begin{align}
\delta_a=
\left\{
\begin{aligned}
&1~~\mathrm{if}~a=0\\
&0~~\mathrm{otherwise}
\end{aligned}
\right.
\end{align}
It is straightforward to check that $A$ and $\mathcal{T}$ satisfy condition (\ref{spin-1/2}), hence $A$ and $\mathcal{T}$ are generators of the symmetry group $\mathbb{Z}_4\rtimes\mathbb{Z}_2^T$ for spin-1/2 fermions. One can also calculate that the two invariants defined above are given by 
\begin{align}
\begin{aligned}
&\mathcal{I}_1=(-1)^{\omega_2(A^2\mathcal{T},A^2\mathcal{T})}=1 \\
& \mathcal{I}_2=(-1)^{\omega_2(A^3\mathcal{T},A^3\mathcal{T})}=-1.
\end{aligned}
\label{app_w2_inv}
\end{align}


\providecommand{\noopsort}[1]{}\providecommand{\singleletter}[1]{#1}%

\end{document}